\documentclass[twocolumn,showpacs,preprintnumbers,amsmath,amssymb]{revtex4}
\usepackage{graphicx}
\usepackage{dcolumn}
\usepackage{bm}
\def\bea {\begin{eqnarray}}
\def\eea {\end{eqnarray}}

\def\be {\begin{equation}}
\def\ee {\end{equation}}
\def\nn {\nonumber}
\begin{document}

\title{Muon pairs from In+In collision at SPS energy }

\author{Jajati K Nayak$^1$, Jan-e Alam$^1$, Tetsufumi Hirano$^2$, Sourav Sarkar $^1$  
and Bikash Sinha$^1$}

\medskip

\affiliation{$^1$Theory Division, Variable Energy Cyclotron Centre, 1/AF, Bidhan Nagar, 
Kolkata 700 064, India}

\affiliation{$^2$Physics Department, The University of Tokyo, Tokyo 113-0033, Japan}
\date{\today}

\begin{abstract}
NA60 collaboration has extracted the inverse slope 
parameters, $T_{\mathrm eff}$ of the dimuon spectra originating 
from the In+In collisions at $\sqrt{s_{\mathrm {NN}}}=17.3$ GeV
for various invariant mass region. They have observed  that the 
inverse slope parameter as a function of invariant mass of the lepton 
pair drops beyond the $\rho$-peak. In the present work,
first we reproduce the  observed invariant mass 
and transverse momentum spectra of the muon pairs.    
Then  show that the slope parameters extracted from the
transverse momentum distributions for various  invariant mass region windows 
can be explained 
by assuming formation of a partonic phase initially which reverts
to hadronic phase through a weak first order 
phase transition at a temperature $T_c\sim 175$ MeV. 
It is observed that a scenario without the formation
of a partonic phase does not reproduce the non-monotonic behaviour
of the inverse slope parameter non typical of radial flow.

\end{abstract}

\maketitle

\section{Introduction}
Low mass lepton pairs are considered as useful tools
to probe the thermodynamic state of the matter presumably
formed in nuclear collisions at relativistic
energies~\cite{mclerran,gale,weldon} (see~\cite{alam1,alam2,rw} for review). 
Unlike hadrons which bring the information of the 
state of the matter when the system is too dilute to support 
any collective behaviour - the lepton pairs probe
the entire stage of the evolution history - 
from the initial formation time to the
final freeze-out time of the fireball.
In this context critical analysis of the high quality muon
data -  available for both the kinematical variables - the transverse
momentum ($p_T$) and the invariant mass ($M$) of the pairs  
from NA60 collaboration~\cite{na60}  assumes prominence
and ignited intense theoretical activities~\cite{hees,ruppert,renk,dusling}. 
Few important observations may be made from the
analysis of the experimental data: the enhancement of lepton pair
productions at the low $M$ ($M < m_\rho$) region indicates
substantial modification of $\rho$ spectral function in the medium; 
noticeable amount of thermal radiation from partonic phase for $M$
beyond $\phi$ peak and finally a non-monotonic behaviour
of the inverse slope parameter, $T_{\mathrm {eff}}$, extracted from
the transverse mass spectra of the pairs - as a function of invariant 
mass displaying development of collectivity 
in the system.  Such a trend may originate due to the
radiation of lepton pairs from a partonic phase formed
initially in the collisions with small average radial velocity.
These considerations call for a simultaneous analysis of the
experimental data as functions of both $M$ and $p_T$. 

The probability that a muon pair of invariant mass, $M$ 
and transverse momentum, $p_T$ 
will emit from a thermal system at temperature $T$ is determined
by  the Boltzmann factor $\sim \exp(-\sqrt{M^2+p_T^2}/T)$ at mid-rapidity.
For a dynamically evolving system like the one produced after
nuclear collisions at ultra-relativistic energies -  the temperature
decreases with time because a part of the thermal energy 
is spent to allow the collective motion of the system. Consequently 
the Boltzmann factor is modified to 
$\sim \exp(-\sqrt{M^2+p_T^2}/T_{\mathrm {eff}})$,
where $T_{\mathrm {eff}}\sim T_{\mathrm {th}}+Mv_r^2$ - here the first term 
represent thermal part and the second term stands for the flow part
with collective radial velocity, $v_r$. 
It is expected that the large $M$
thermal pairs will be emitted from early time when temperature is large and 
(radial) flow velocity is small
and the small $M$ pairs will originate from the late stage of the
evolution when the temperature is low but transverse flow velocity is large.  
Therefore, a simultaneous study of the dilepton spectrum as functions of
$p_T$ and $M$ will enable us to use the
variation of the $T_{\mathrm {eff}}$ with $M$ 
as a chronometer of the heavy ion collisions.

Accordingly in the present  work,  we focus on  the transverse momentum
and the invariant mass distributions
of the lepton pairs from in In + In collisions
at 158A GeV beam energy at CERN-SPS.  
We assume the following two scenarios for the collisions: 
(i)In+In $\rightarrow$ quark gluon plasma (QGP) $\rightarrow$ mixed phase of 
quarks and hadrons $\rightarrow$ hadronic phase and (ii) In+In $\rightarrow$ 
hadronic phase and check by comparing the results with 
$p_T$ distribution of dileptons which is the possible  
scenario realized in the collisions. 

The paper is organized as follows. 
In the next section we present the dilepton productions from
QGP and hadronic matter. Section III is dedicated to the initial
conditions and the space time evolution. Section IV contains the
results and section V is devoted to summary and discussions.

\section{Dilepton productions from hadronic matter and quark gluon plasma}
The rate of thermal dilepton production per unit space-time volume
per unit four momentum volume is given by\cite{mclerran,gale,weldon}
\be
\frac{dR}{d^4p}=-\frac{\alpha^2}{6\pi^3q^2}L(M^2)f_{\mathrm{BE}}(q_0)
{W_\mu^{\mu}}(q_0,\vec q)
\label{virtual}
\ee
where $\alpha$ is the electromagnetic 
coupling, $W_\mu^{\mu}$ is the correlator of electromagnetic currents
and $f_{\mathrm {BE}}(E,T)$ is the thermal phase space factor for Bosons.
The factor
\be
L(M^2)=\left(1+2\frac{m^2}{M^2}\right)\sqrt{1-4\frac{m^2}{M^2}}
\label{spinor}
\ee
arises from the final state leptonic current involving Dirac
spinors of mass $m$ (in this case muon) and
$p^2(=p_\mu p^\mu)=M^2$ is the invariant mass square of the lepton pair.
In hadronic matter, this can be simplified using vector meson dominance
to give (see~\cite{Sabya_dil} for details),
\be
\frac{dR}{dM^2q_Tdq_Tdy}=\frac{\alpha^2}{\pi^2M^2}L(M^2)f_{\mathrm{BE}}(q_0)
\sum_{V=\rho,\omega,\phi}A_V(q_0,\vec q)
\label{dil2}
\ee
where the spectral function of the vector mesons consists of a pole and
continuum,
\be
A_V=A_V^{\rm pole}+A_V^{\rm cont}
\ee
For the $\rho$, the continuum part is parametrized as~\cite{alam2,shuryak}
\be
A_\rho^{\rm cont}=\frac{m_\rho^2}{8\pi}\left(1+\frac{\alpha_s}{\pi}\right)
\frac{1}{1+\exp(\omega_0-q_0)/\delta}
\ee
with $\omega_0=1.3$ GeV and $\delta=0.2$ GeV and the pole part is given
by~\cite{Sabya_dil}
\bea
A_\rho^{\rm pole}&=&-\frac{f_\rho^2m_\rho^2}{3}\left[\frac{2\sum{\rm
Im}\Pi^R_t}{(q^2-m_\rho^2-\sum\mathrm{Re}\Pi^R_t)^2
+(\sum{\rm Im}\Pi^{R}_t)^2}\right.\nonumber\\&&\left.+\frac{q^2\sum{\rm Im}\Pi^R_l}
{(q^2-m_\rho^2-q^2\sum\mathrm{Re}\Pi^R_l)^2
+q^4(\sum{\rm Im}\Pi^{R}_l)^2}\right]
\label{eq:spdef}
\eea
with $f_\rho=0.130$ GeV. 
As we have included the continuum in the vector
mesons spectral functions four pion annihilation process~\cite{4pi}
is not considered here to avoid over counting.
The self-energy $\Pi$ contains contributions from
mesons as well as baryons in the thermal medium so that 
\be
\Pi=\Pi_M+\Pi_B
\ee
The longitudinal and transverse components of the meson part $\Pi_M$ has been
evaluated in detail for one loop $\pi-h$ graphs with
$h=\pi,\omega,h_1,a_1$ in the real time formulation of thermal field theory
~\cite{Sabya_EPJC}. The
baryonic contribution $\Pi_B$ has been estimated in the approach of Eletsky   
et al~\cite{Eletsky} using resonance dominance in the low energy region
and a Regge model at higher energies.
Dilepton emission from the $\omega$ and the $\phi$ have 
also been included. The
width of the $\omega$ in thermal bath is taken from the calculation of
Ref~\cite{Weise} where a framework similar to the one employed here has
been used. For the $\phi$ only the vacuum width has been considered.

The major source of lepton pair production
is due to the annihilation of quark-antiquark pairs~\cite{cleymans}.
In the present calculations the QCD corrections through the 
processes: 
$q\bar{q}\rightarrow g l^+l^-$,\,
$gq(\bar{q})\rightarrow q (\bar{q})l^+l^-$ (see Ref.~\cite{altherr,thoma}
for details) in the dilepton productions have also 
been taken into account.

To evaluate the thermal dilepton spectra the static (fixed
temperature) rate of emission
($dR/d^2m_TdM^2dy$) has to be convoluted with the space-time
dynamics governed by the relativistic hydrodynamics as follows:

\bea
\frac{dN}{m_{T}dm_{T}}&=&2\pi \sum_{phases}\int
{\left(\frac{dR}{d^2m_{T}dydM^2}\right)_{phase}}\nn\\
&&\times dM^2 dyd^4x
\label{eq8}
\eea
$d^4x$ is the 4 dimensional volume element,
$m_T$ is the transverse mass and $y$ is the rapidity. 
The limits for integration over invariant mass, $M$ can be fixed according to 
the experimental measurements. For the experimental data for
$M_{\mathrm {min}}\leq M \leq M_{\mathrm {max}}$, the 
transverse mass $m_T$ is defined
as $m_T=\sqrt{M^2+p_T^2}$, where $M=(M_{\mathrm {min}} 
+M_{\mathrm {max}})/2$.   The invariant mass
spectra of the lepton pairs can be obtained from Eq.~\ref{eq8}
by integrating over the appropriate $p_T$ windows. The acceptance 
corrections due to detector geometry has been taken into consideration.

We need to also consider the contributions from $\rho$-decay as
this has not been subtracted from the data (see second of ~\cite{na60}).
To evaluate the $m_T$ distribution of dilepton  
originating from the decays of vector mesons after freeze-out 
the  Cooper-Frye formula~\cite{cooper} has been  used.
For a special case of unstable vector mesons  
we need to know the thermal phase space factor 
corresponding to  an unstable boson which is given by 
\begin{equation}
f_{\mathrm {unstable}}=
\frac{g}{(2\pi)^3}\,\frac{1}{\mathrm {exp(p_0/T)} - 1}\rho(M)
\end{equation}
where $p_0=\sqrt{p^2+M^2}$, $g$ is the statistical degeneracy and 
$\rho(M)$ is the spectral function of the vector meson under 
consideration. For stable particle $\rho(M)$ reduces to a Dirac
delta function and consequently the usual thermal phase space factor for
a stable particle is  recovered. 
Therefore, the $m_T$ distribution of dimuons from vector meson decay after the
freeze-out is given by
\begin{figure}
\begin{center}
\includegraphics[scale=0.45]{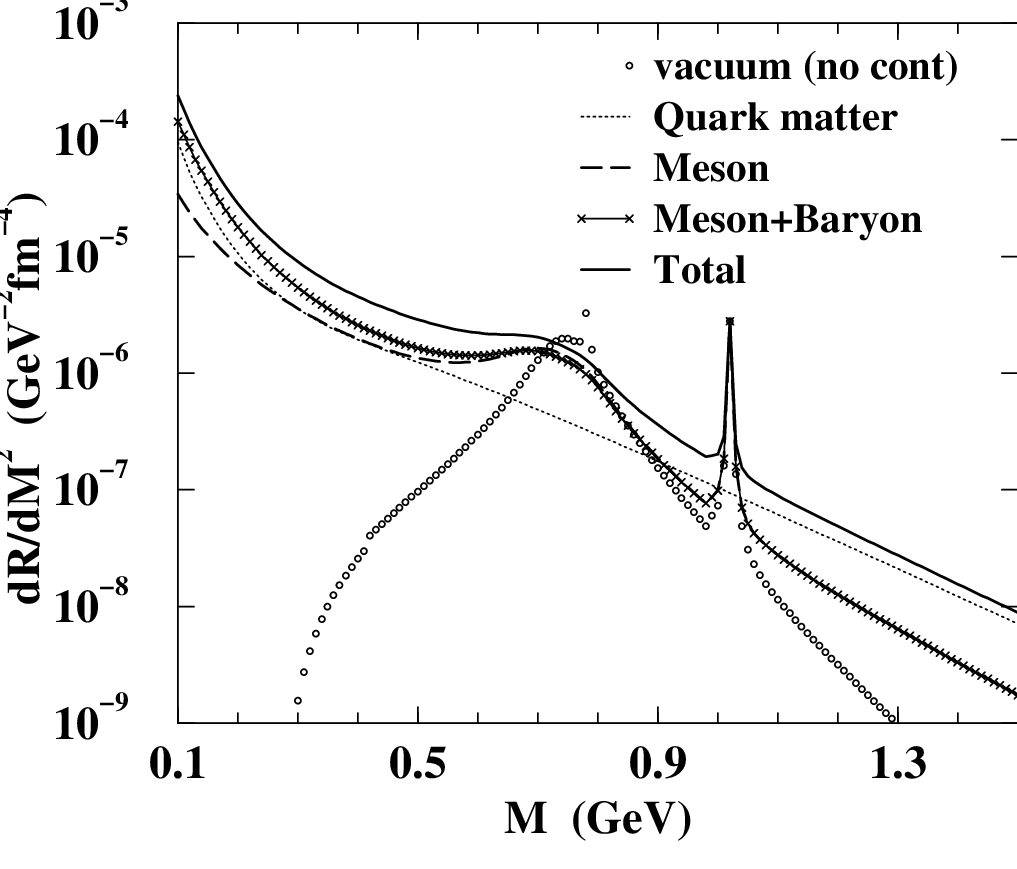}
\caption{Invariant mass spectra (acceptance corrected inclusive mass spectrum
$p_T >0$) of dimuon from quark gluon plasma and hadronic phase at $T=175$ MeV
and $\mu_B=250$ MeV.
}
\label{fig1}
\end{center}
\end{figure} 

\bea
\frac{dN_{\gamma^\ast}}{m_{T}dm_{T}}&=&2\pi\int dr \int d\eta \int d\phi\, 
r\tau\nn\\
&&\times \left(m_T\mathrm {cosh}(y-\eta)-\frac{\partial\tau}{\partial r}p_T
\mathrm {cos}\phi\right)\nn\\
&&\times \rho(M)\Gamma_{V\rightarrow \mu^+\mu^-}/\Gamma_V^{\mathrm {tot}}
f_{\mathrm {unstable}}\nn\\ 
&& dM^2dy
\label{f.o.}
\eea
where $\Gamma_V^{\mathrm {tot}}$ is the total decay width of the vector meson, $V$.
The contributions from Eqs.~\ref{eq8} and \ref{f.o.} are added  for the
description of the  $m_T$ spectra of the data. 

The other non-thermal sources are:
(i) Drell-Yan process originating from the interactions of quarks
and anti-quarks of the colliding nuclei and the  decays of various 
mesons ({\it e.g.} $\pi$,$\eta$, $\omega$, $\rho$, $\eta^\prime$, $\phi$ etc)
after the fireball freeze-out. As the non-thermal contributions 
have been subtracted from the data under consideration,
we concentrate only on the thermal emissions.

\section{Initial conditions and the space time evolution}
The space time evolution of the system 
has been studied using ideal relativistic hydrodynamics 
with longitudinal boost 
invariance ~\cite{bjorken} and cylindrical symmetry~\cite{von}. 
The initial temperature($T_i$) 
and thermalization time ($\tau_i$) are constrained 
by the following equation ~\cite{Hwa} for an isentropic expansion:
\be
T_i^{3}\tau_i \approx \frac{2\pi^4}{45\xi(3)}
\frac{1}{4a_{\mathrm{eff}}}\frac{1}{\pi R_A^2}\frac{dN}{dy}.
\label{eq6}
\ee
where, $dN/dy$= hadron multiplicity,  
$R_A$ is the effective radius of the system (evaluated by
using the formula, $R_A\sim N_{part}^{1/3}$), 
$\xi(3)$ is the Riemann zeta function and  $a=\pi^2g/90$
($g=32$, taken as the effective degeneracy of the QGP phase). 
The initial radial velocity, $v_r(\tau_i,r)$ 
 and energy density, $\epsilon(\tau_i,r$) 
profiles are taken as:
\be
v_r(\tau_i,r)=0,\,\,\,\,\,
\epsilon(\tau_i,r)={\epsilon_0}/({e^{\frac{r-R_A}{\delta}}+1}) 
\label{eq7}
\ee
where the surface thickness, $\delta=0.5$ fm. 
In the present work we assume $T_c = 175$ MeV~\cite{borsanyi}. 
In a quark gluon plasma to hadronic transition scenario - 
we use the bag model EOS for the QGP phase and for the hadronic 
phase all the resonances with mass $\leq 2.5$ GeV have been 
considered~\cite{bm}.  
The transition region has been parametrized as follows~\cite{hatsuda}:
\be
s=f(T)s_q + (1-f(T))s_h
\ee
where $s_q$ ($s_h$) is the entropy density of the quark (hadronic) 
phase at $T_c$ and 
\be
f(T)=\frac{1}{2}(1+\mathrm {tanh}(\frac{T-T_c}{\Gamma}))
\label{eos}
\ee
the value of the parameter $\Gamma$ can be varied to
make the transition strong first order or continuous. 
\begin{figure}
\begin{center}
\includegraphics[scale=0.45]{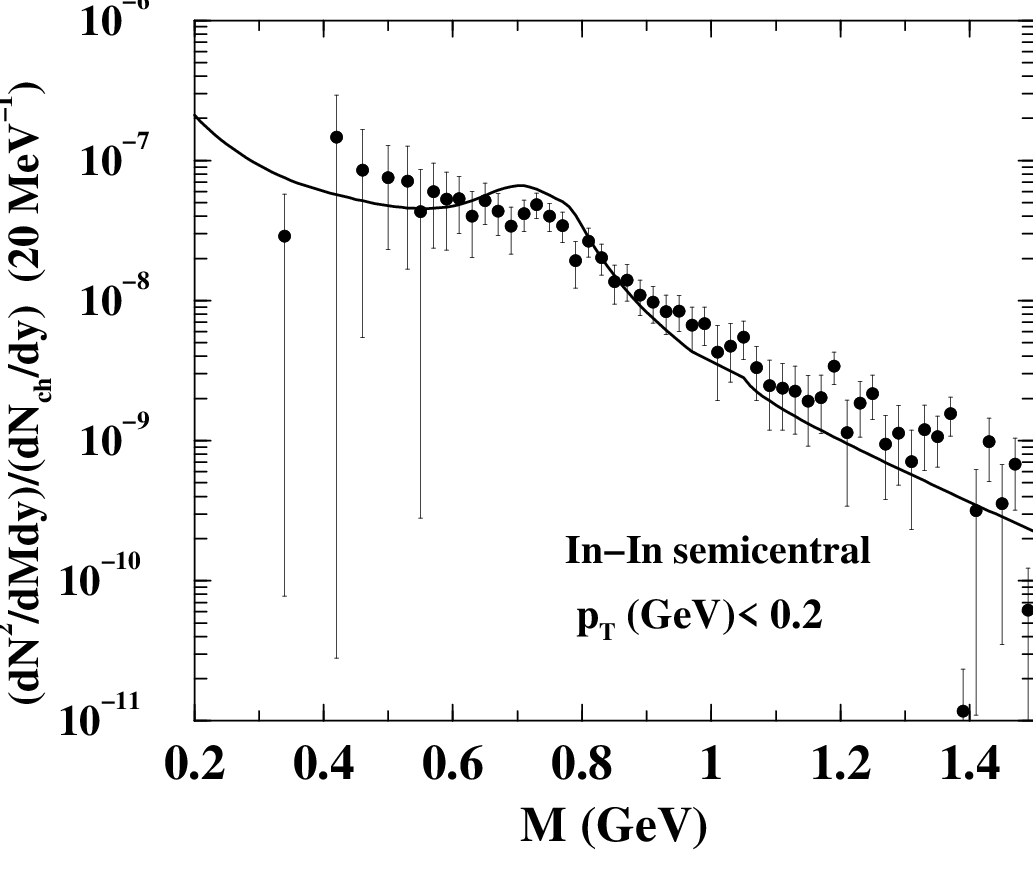}
\includegraphics[scale=0.45]{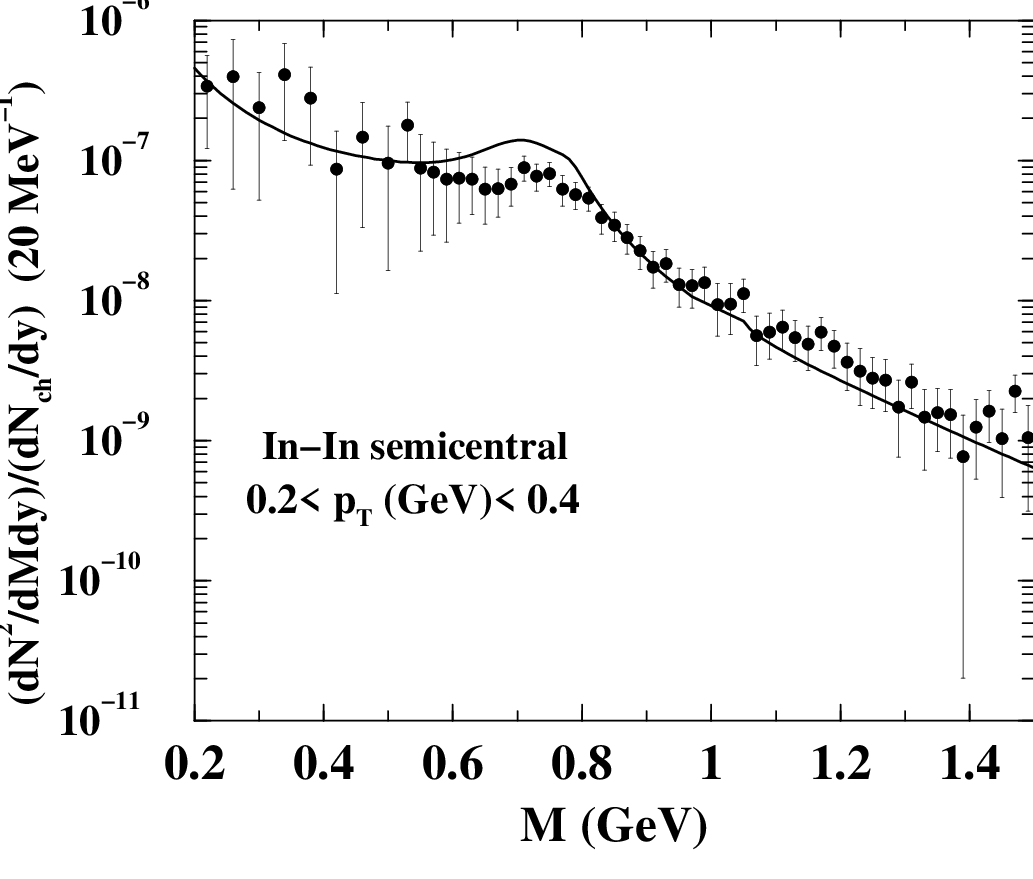}
\caption{Invariant mass spectra (acceptance corrected inclusive mass spectrum)
for different $p_T $ window of the dimuon measured by NA60 collaboration for
semi central In+In collision ($\sqrt{s_{NN}}=17.3$ GeV). The solid line is the
theoretical result.
}
\label{fig2}
\end{center}
\end{figure}
\begin{figure}
\begin{center}
\includegraphics[scale=0.45]{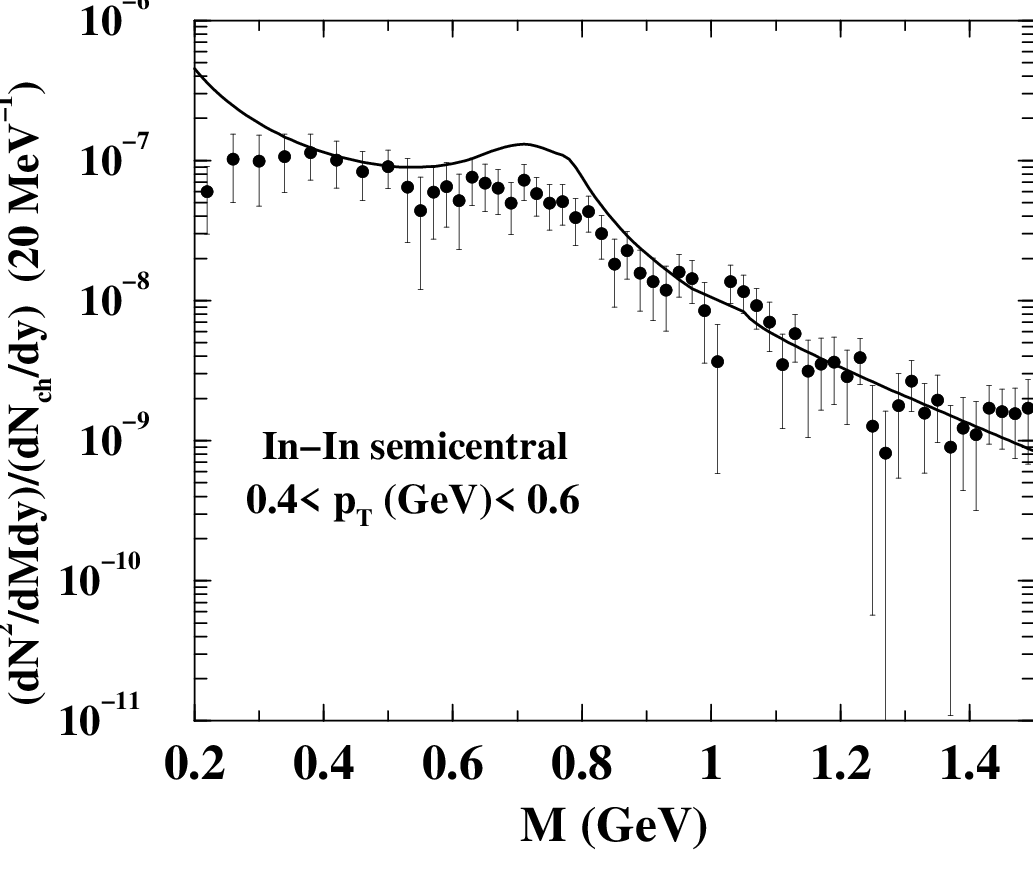}
\includegraphics[scale=0.45]{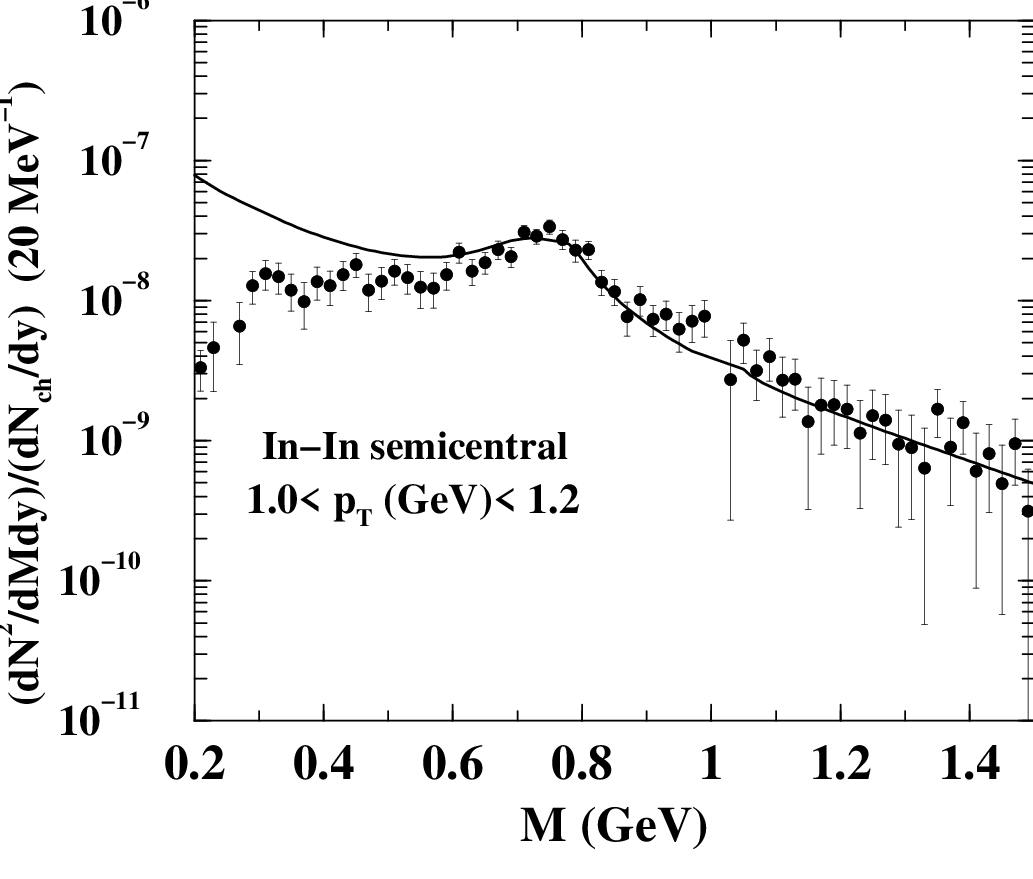}
\caption{Invariant mass spectra (acceptance corrected inclusive mass spectrum)
for different $p_T $ window of the dimuon measured by NA60 collaboration for
semi central In+In collision ($\sqrt{s_{NN}}=17.3$ GeV). The solid line is the
theoretical result.
}
\label{fig3}
\end{center}
\end{figure}

The ratios of various hadrons measured experimentally at 
different $\sqrt{s_{\mathrm {NN}}}$
indicate that the system formed in heavy ion collisions chemically decouple 
at a temperature ($T_{\mathrm {ch}}$) which is higher than the temperature for 
kinetic freeze-out ($T_f$)determined
by the transverse spectra of hadrons~\cite{pbm}. 
Therefore, the system remains out of chemical equilibrium
from $T_{\mathrm {ch}}$ to $T_f$.
The chemical non-equilibration affects the dilepton yields  at two levels: 
a) the emission rate through the phase space  factor and 
b) the space-time evolution
of the matter through the equation state. 
The value of the chemical potential and
its inclusion in the EOS has been taken in to account 
following Ref.~\cite{hirano}. It is expected that the chemical
potentials do not change much for the inclusion of resonances
above $\Delta$.

\begin{figure}
\begin{center}
\includegraphics[scale=0.45]{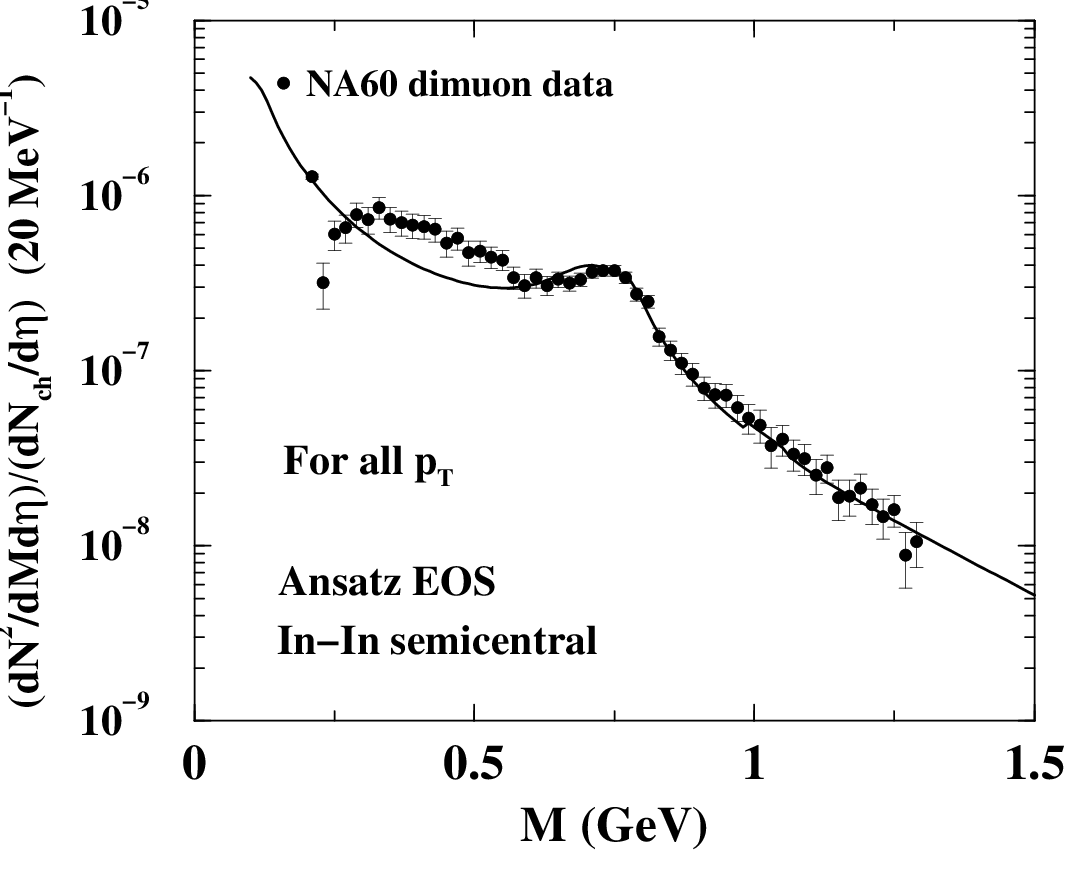}
\caption{Invariant mass spectra (acceptance corrected inclusive mass 
spectrum $p_T >0$) of dimuon measured by NA60 collaboration for semi central 
In+In collision ($\sqrt{s_{NN}}=17.3$ GeV). The solid line is the theoretical 
result.
} 
\label{fig4}
\end{center}
\end{figure} 

\section{Results}
Before discussing the $m_T$ spectra it is important
to mention here that initially there was no transverse collective
flow ($v(\tau_i,r)=0$), the entire energy of the system
was thermal. With the progress of time some
part of the thermal energy gets converted to the collective (flow) energy
for a system undergoing hydrodynamic expansion. 
The measured $m_T$ spectra of muon pairs therefore, contains contributions 
from both thermal as well as collective degrees of freedom {\it i.e.} 
the inverse slope parameter, $T_{\mathrm {eff}}$ can be written as 
$T_{\mathrm {eff}}=
T_{\mathrm {th}} + Mv_r^2$ as mentioned earlier. Therefore, it is important
to know the domains of $M$ where the thermal contributions from early
(quark matter) and late (hadronic) phases dominate corresponding
to small and large radial flow respectively.  

We discuss results for scenario (i) first - here we have
assumed that a thermalized state of quarks and gluons is
formed after the collisions which reverts to hadronic phase
through a weak first order phase transition. The
value of $\Gamma$ in Eq.~\ref{eos} is taken as 20 MeV. 
The values of initial energy density, $\epsilon_i=5.5$ GeV/$fm^3$, 
thermalization time $\tau_i=0.6$ fm
and the transition temperature 
$T_c=175$ MeV have been assumed. 
We take the freeze-out temperature 
$T_f=130$ MeV which can reproduce the slope of
the $\phi$ spectra measured by NA60 collaboration for
In+In collisions~\cite{floris}. 
With these inputs transverse and invariant 
mass spectra of dileptons have been evaluated.

We look into the $M$ spectra. 
A significant
enhancement in the dilepton yield in the mass region below the $\rho$ pole
(compared to vacuum, denoted by dots in Fig.~\ref{fig1})
is observed, nevertheless the total dilepton yield in this
region of $M$ contains notable contribution from the partonic phase
(~Fig.\ref{fig1}).
However, the thermal pairs 
for $M$ beyond $m_\phi$-peak is dominated  by QGP phase. 
Therefore, it is expected  
that the slope parameters extracted from the transverse mass 
distribution of lepton pairs for mass region above the
the $\phi$-peaks  would reflect the properties of
quark matter phase, {\it i.e.}
the  slopes at these $M$ region will correspond
to the early time when the radial flow is small.
On the other hand the contributions in the region of
$\rho$ mass  are overwhelmingly from the hadronic phase
and hence the slope at this region correspond to the
late time containing large radial flow effects.
For $M<m_\rho$ the situation is complex as it contains significant 
contributions from both the hadronic as well as the partonic 
phase.

In Figs.~\ref{fig2}-\ref{fig3} the invariant mass
spectra for various $p_T$ windows have been displayed.  
At low $p_T$ and low 
$M$ a clear rise observed in the data - reflecting the thermal nature
of the radiation. 
The strong enhancement at the low $M$ domain is reproduced
well due to large broadening of the $\rho$ in the thermal medium for
the $p_T$ (in GeV) windows:
$p_T<0.2$, $0.2<p_T<0.4$ and $0.4<p_T<0.6$. 
For the higher $p_T$ window ($1.0<p_T<1.2$), the observed plateau
for $0.3<M$(GeV)$<0.6$ is over  estimated  by the theoretical yield, 
which is yet to be understood~\cite{jknfuture}.
The invariant mass spectra, without any $p_T$ selection is reproduced
reasonably well (Fig.~\ref{fig4}). The high $M$ region, above $\phi$ peak,
the radiation from the partonic phase describe the data
very well.

\begin{figure}
\begin{center}
\includegraphics[scale=0.45]{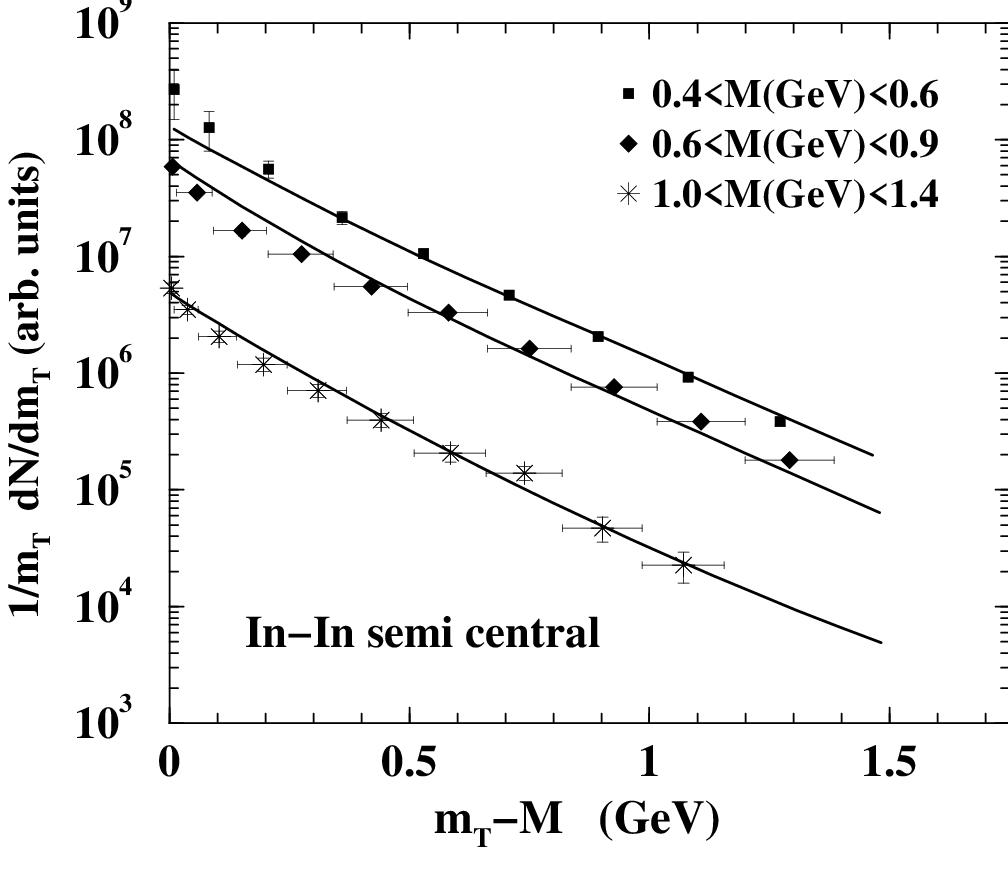}
\caption{$m_{T}-M$ spectra of muon pair for different invariant mass ranges
for semi central In-In Collision at $\sqrt{s_{NN}}=17.3$ GeV (158 A GeV).
}
\label{fig5}
\end{center}
\end{figure}

The resulting $m_T-M$ spectra of 
$\mu^{+}\mu^{-}$ are compared with the experimental data obtained 
by the  NA60 collaboration for In-In collision 
for $\sqrt{s_{\mathrm {NN}}}=17.3$ GeV
for different mass window in Fig.~\ref{fig5}. 
The present calculation agrees well with the data which is shown 
by solid lines for all the mass ranges. 

\begin{figure}
\begin{center}
\includegraphics[scale=0.45]{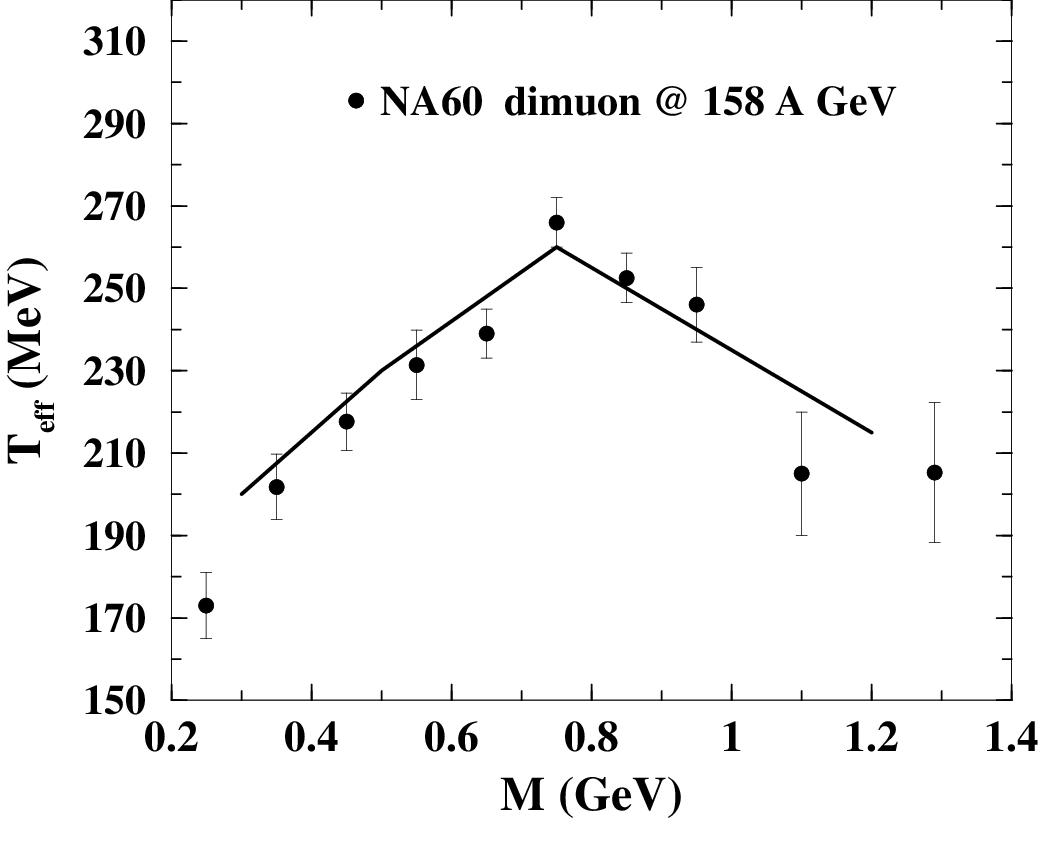}
\caption{Inverse slope parameter obtained from the $m_{T}-M$ spectra
is plotted with the average invariant mass}. 
\label{fig6}
\end{center}
\end{figure}

In Fig~\ref{fig6} the effective temperature obtained from the inverse 
slope of these spectra 
has been  plotted and compared with the data. 
The slopes have been estimated from theoretical results (shown
by solid lines in Figs.~\ref{fig5}) by parameterizing
to an exponential function 
within the $(m_T-M)$ range $0.3\leq m_T-M$(GeV)$\leq 1.0$. 
It is clear from the
results that the slope at high $M$ region is reproduced well if 
the source is predominantly partonic.
A similar non-monotonic behaviour is observed in the variation of the elliptic
flow ($v_2$) of photons as a of transverse momentum~\cite{rupa,liu}. 
 
\begin{figure}
\begin{center}
\includegraphics[scale=0.45]{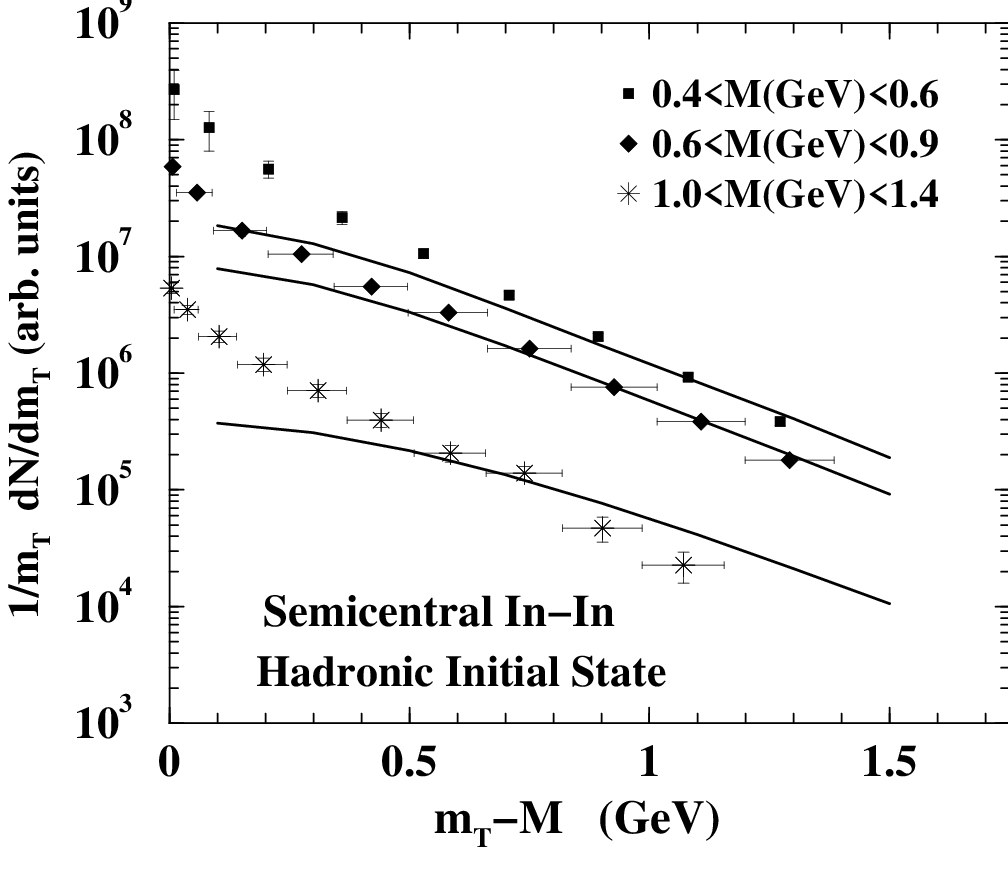}
\caption{$m_{T}-M$ spectra of muon pair for different invariant mass ranges
for semi central In-In Collision at $\sqrt{s_{NN}}=17.3$ GeV (158 A GeV).
for hadronic initial state {\it i.e.,} scenario (ii).
}
\label{fig7}
\end{center}
\end{figure} 
Next we discuss results for scenario (ii). In this scenario
we evolve a hadronic matter with same initial energy density 
$\epsilon_i$=5.5 GeV/$fm^3$ up to freeze-out temperature 
$T_f=130$ MeV with an equation of state which includes 
mesons and baryons up to mass 2.5 GeV. 
The equation of state obtained for such a scenario is soft
compared to an initial QGP state, the average
velocity of sound ($c_s$) is 
$c_s^2=0.16$. With this scenario we fail to reproduce the $m_T$
distribution (Fig.~\ref{fig7}),
the inverse slope parameter extracted in this scenario increases monotonically
with $M$ (not shown in Fig.~\ref{fig6}, but is obvious that 
the slopes extracted from Fig.~\ref{fig7} will fail to reproduce
the data in Fig.~\ref{fig6}) .

We have checked that a large increase in $T_c$ leads to a
poorer description of the data.  
In this case the contributions
from the QGP phase is considerably reduced so the 
slopes of the $m_T$ spectra is pre-dominantly determined 
by the hadronic phase.
In this scenario the slopes for $M$ beyond the $\phi$ peak is  not
determined by the (early) QGP phase but by the (late) hadronic
phase.
As a result the sharp fall
in the   $T_{\mathrm {eff}}$ as a function of $M$ for $M>m_\rho$
is not properly reproduced. This
indicates that the source of lepton pairs at large 
($M\sim 1.2$ GeV) is partonic.

\section{Summary and discussions}
In summary we have reproduced  the invariant mass and transverse
momentum spectra of the lepton pairs measured by NA60 
collaboration in In+In collisions
at $\sqrt{s_{\mathrm NN}}=17.3$ GeV. 
The broadening of the $\rho$ spectral function due to
its interaction with mesons and baryons in the thermal bath
has been taken into account. The study reveals that the 
description of the data 
at low $M$ needs large broadening of $\rho$ meson and
at high $M$ (beyond $\phi$ peak) requires substantial contributions
from partonic phase. 
We find that the  measured slope 
of the dilepton spectra  can be
explained theoretically if one assumes the formation 
of a partonic phase initially which reverts to hadronic
phase through a weak first order phase
transition. A hadronic initial state with realistic equation 
of state fails to reproduce the data.

\par 
{\bf Acknowledgment:}
We are grateful to Sanja Damjanovic and Hans Specht for providing us the 
experimental data and detector acceptance 
and having a fruitful discussion. 
We would like to thank Bedangadas Mohanty and Nu Xu for helpful 
discussions.  JA is  supported by DAE-BRNS
project Sanction No.  2005/21/5-BRNS/2455.
\par 

\end{document}